\documentclass{dune}
\pdfoutput=1

\input{common/preamble}

\begin{document}
%\linenumbers
\pagestyle{titlepage}
%\cleardoublepage

% This should be \input first thing after \begin{document}

\pagestyle{titlepage}

%\title{\scshape\Large Snowmass Neutrino Frontier: \\
%NF07 Topical Group Report\\
%\vspace{5mm}
%\normalsize Applications

\title{\scshape\Large Report of the Topical Group on Neutrino Applications for Snowmass 2021

\normalsize
\vskip -10pt
\snowmasstitle
}

%\vspace{10mm}

\renewcommand\Authand{, }%<---------------remove and
\renewcommand\Authands{,}
\renewcommand\Authfont{\scshape\small}
\renewcommand\Affilfont{\itshape\footnotesize}

%Authors go here with affiliation indexed by [X]
\author[1]{N.~S.~Bowden}
\author[2]{J.~M.~Link}
\author[3]{W.~Wang}
\author[ ]{\\on behalf of the NF07 Applications Topical Group Community}

\vspace{-0.5cm}
%Affiliations go here
\affil[1]{Lawrence Livermore National Laboratory, Livermore, CA 94550, USA}
\affil[2]{Center for Neutrino Physics, Virginia~Tech, Blacksburg, Virginia  24061, USA}
\affil[3]{Sun Yat-Sen (Zhongshan) University, Guangzhou, China}

\date{}

\maketitle

\renewcommand{\familydefault}{\sfdefault}
\renewcommand{\thepage}{\roman{page}}
\setcounter{page}{0}

\pagestyle{plain} 
\clearpage
\textsf{\tableofcontents}

%\textsf{\listoffigures}
%\clearpage

%\textsf{\listoftables}
%\clearpage

%For acronym list to appear just after TOC, TOF, TOT
%\printnomenclature
%\clearpage

%\iffinal\else
%\textsf{\listoftodos}
%\clearpage
%\fi

\renewcommand{\thepage}{\arabic{page}}
\setcounter{page}{1}

\pagestyle{fancy}

% Set how header/footers look
%\newcommand{\chaptermark}[1]{%
%\markboth{Chapter \thechapter:\# 1}{}}
%\renewcommand{\chaptermark}[1]{%
%\markboth{Chapter \thechapter:\ #1}{}}
\fancyhead{}
%\fancyhead[RO,LE]{\textsf{\footnotesize \thepage}}
\fancyhead[RO]{\textsf{\footnotesize \thepage}}
%\fancyhead[LO,RE]{\textsf{\footnotesize \rightmark}}
\fancyhead[LO]{\textsf{\footnotesize \rightmark}}

\fancyfoot{}
\fancyfoot[RO]{\textsf{\footnotesize Snowmass 2021}}
\fancyfoot[LO]{\textsf{\footnotesize NF07 Topical Group Report}}
\fancypagestyle{plain}{}

\renewcommand{\headrule}{\vspace{-4mm}\color[gray]{0.5}{\rule{\headwidth}{0.5pt}}}

% Not all main documents have any citations.
% When not built in "final" mode, add in one citation just to let the
% document build.
% If, after substantial editing a main document still lacks any
% citations then it should have its whole bibliography removed.
%\ifdefined\isfinal\nocite{}\else\nocite{CD0}\fi
%\nocite{nothing}

% see also preamble.tex
%\input{common/acronyms}

\clearpage

\section*{Executive Summary}
\label{sec:summary}
\addcontentsline{toc}{section}{Executive Summary}

In addition to the pursuit of fundamental knowledge, investment in scientific discovery is also motivated by the development of technologies and the training of a skilled workforce that can benefit society in other ways. In the case of Neutrino Physics, there are many instances of this dynamic playing out. But somewhat surprisingly, given the inherent difficulty of neutrino detection, the direct application of neutrinos to advance other fields of research or solve societal problems is also a possibility. Examples of direct neutrino applications include monitoring of nuclear reactors for safeguards and nonproliferation and the probing the Earth's interior via it's neutrino emissions and tomography.

There are strong synergies between the neutrino physics topics of community interest developed through this Snowmass process and direct neutrino applications. These synergies can take the form of overlapping needs in terms of technology development, workforce capabilities, facilities, and underlying scientific knowledge. Specific examples of synergies highlighted during Snowmass include:
\begin{itemize}
    \item Improved knowledge of the reactor neutrino emissions (flux and spectrum) is needed for reactor monitoring applications and would enable reactors neutrinos to continue to be a tool for discovery;
    \item Neutrino detectors using the inverse beta decay or CEvNS channels with improved background suppression operating close to reactors can help to address neutrino anomalies, probe BSM physics, and provide versatile monitoring capabilities;
    \item Technologies that could reduce the cost and improve the performance of large inverse beta decay neutrino detectors could provide reactor monitoring at larger distances and address a wide range of neutrino physics topics; 
    \item Projects at the intersection of neutrino physics and neutrino applications can provide broad training  opportunities for community members while also opening new career pathways.    
 \end{itemize}

The fact that technologies and workforce capabilities developed by the field of Neutrino Physics can have benefits in many other areas of science and society provides an additional motivation for societal investmen. 
%in the field and pursue the scientific goals of the community. 
To further advance our field in this respect, it is important that community members have such possibilities in mind and actively seek opportunities to highlight the broader impact of their research. It is also critical to note that the full impact of neutrino applications can only be realized through early and attentive engagement with experts from potential end user communities. 

Given the strong synergies between neutrino physics and neutrino applications noted above, it is apparent that stakeholders from both communities would benefit from further coordination. This could take the form of joint investment in detector R\&D efforts, measurements to improve our knowledge of reactor emissions, and/or reactor-based experiments and demonstrations.

% [\textbf{Pending update:} Exec summary narrative text]

% Key Takeaways
% \begin{itemize}
%     \item The technologies and knowledge created by neutrino physicists have utility beyond our field through the application of technologies and the direct use of neutrinos themselves.
%     \item Neutrino physics and direct neutrino applications require  strongly overlapping technology and workforce capabilities.
%     \item There are strong synergies between the future scientific goals, nuclear data needs, and technology pathways of neutrino physics and direct neutrino applications.
%     \item Stakeholders for both fundamental and applied neutrino physics programs would benefit from coordination of investments in reactor-based experiments and demonstrations.
% \end{itemize}
% %%%%%%%%%%%%%%%%%%%%%%%%%%%%%%%%%%%%%%%%%%%%%%%%%%%%%%%%%%%

\cleardoublepage

\section{Introduction and Motivation}
\label{sec:introduction}

The pursuit of science is a great endeavour, that in addition to creating new fundamental knowledge, improves our well-being through education and by enabling new technologies~\cite{UNESCO}. In this Topical Group Report on Applications of Neutrino Physics, we explore these latter two functions of our collective endeavour. Here, we will explore and develop the overlap between applications and active fields of research within the neutrino community, while also highlighting areas where stronger connections could be developed with other disciplines, user communities, and sponsor agencies. 

Since the technology and skills that neutrino physicists develop are often pushing the boundaries of what can be achieved, it is natural that they might also solve problems beyond our specific discipline. 
Furthermore, as our knowledge of the underlying neutrino physics continues to improve in concert with detector technology, the door has opened to using the unique nature of the neutrino to access information where it is otherwise difficult or impossible. 

This report is structured as follows. Section~\ref{sec:applications} begins with descriptions of several areas where Neutrino Physics and it's supporting technologies find actual or potential application to other areas. Particular emphasis is given to the possible direct use of neutrinos to solve problems in Nuclear Security and Nuclear Energy. An active sub-field has grown around the idea using  electron antineutrinos released as a by-product of nuclear fission for this purpose.  Section~\ref{sec:applications} concludes with a  description of potential capabilities conceived primarily by neutrino physicists, after which Section~\ref{sec:nuclear-applications} gives the perspective of potential end-users of what would be a rather novel technological approach. 

A central theme of community input and this report is that Neutrino Physics and potential Applications can mutually benefit from continued engagement and investment. Section~\ref{sec:synergies} develops this theme by examining synergies between scientific and application topics. This report concludes in Section~\ref{sec:nextsteps} with an examination on Next Steps that we as a community can take to fully realize these synergies, advance both the fundamental and applied aspects of our field and continue to benefit society in the broadest possible way.

%%%%%%%%%%%%%%%%%%%%%%%%%%%%%%%%%%%%%%%%%%%%%%%%%%%%%%%%%%%
%\cleardoublepage

\section{Applications of Neutrino Physics and Supporting Technologies}
\label{sec:applications}

Applications of Neutrino Physics can be approximately grouped into two categories:  (1) areas that benefit from technologies and skills developed in the pursuit of neutrino physics and (2) areas where neutrinos themselves can be used to address a need, which in turn requires an understanding of neutrino production and neutrino properties, plus the ability to detect neutrinos. In this section we give several examples of category (1), but by no means identify all such  applications. The consistent theme that we wish to highlight with these examples is that the pursuit of new neutrino knowledge requires physicists to develop new and improved technologies, which can also redefine what is possible in other areas. And of course, the converse is also true where advances developed for other purposes open new lines of pursuit in Neutrino Physics.

The second category,  use of neutrinos to address needs in other fields, is a relatively recent development,
%. For the same reason that (someone) declared (insert quote about neutrinos being undetectable..), 
the small size of the weak interaction and the consequent difficulty of neutrino detection limit the opportunities for direct application. Several direct applications have emerged in recent years. The use of `geoneutrinos' to probe the inner structure of the Earth will be covered by the NF04 `Neutrinos from Natural Sources' Topical Group. Neutrinos from a variety of sources could also be used for tomography of the Earths interior~\cite{Winter:2006vg,Donini:2018tsg}. Technically conceivable applications of the electron antineutrinos emitted as a by-product of nuclear fission, {\it e.g.} in reactors, to problems in Nuclear Security and Nuclear Energy was the subject of several community submissions to Snowmass and will be covered here. 

\subsection{Application of Technologies Developed for Neutrino Physics}

\subsubsection{Isotope Extraction and Enrichment}

Planning for the next generation of neutrinoless double beta decay ($0\nu\beta\beta$) experiments beyond the tonscale is already underway. Such experiments would need very large quantities of  $0\nu\beta\beta$ isotopes, which will require new approaches to enrichment in order to be logistically and economically feasible. To take the specific example of xenon, Metal-Organic Frameworks (MOFs) for extracting xenon efficiently from enormous quantities of atmospheric air is one approach that is under development~\cite{Avasthi:2021lgy}. Such technologies could be applied to atmospheric sampling for radio-xenon detection in the context of the Comprehensive Test Ban Treaty (CTBT)~\cite{Auer}. Additionally, xenon use in applications is severely limited by its high cost. Greater availability would have the potential for large societal impact, {\it e.g.} in medicine for anesthesia and medical imaging, and space applications for use in ion thrusters for satellite stationkeeping~\cite{Sangiorgio_PC}. 

\subsubsection{Accelerator Technology}

Cyclotrons can be an important source of radioactive isotopes needed for medical imaging, like $^{68}$Ge, and cancer treatments, like $^{225}$Ac~\cite{Alonso:2022roj}.  The IsoDAR project~\cite{IsoDAR:2021loi} is developing a high powered cyclotron to produce an intense neutron beam from $^8$Li beta decay-at-rest.  Due to its high power, the IsoDAR cyclotron would represent an order of magnitude improvement over the current state of the art in cyclotron-based medical isotope production. 

\subsubsection{Radiation Detection Technology}

There is a long history of detection techniques developed for particle physics being used for more general radiation detection purposes. Advances in materials, DAQ, and detection concepts are all potentially relevant to improving detection capabilities for a wide range of applications like nuclear nonproliferation, homeland security, and industry. Particle physicists are an important workforce pipeline into these fields. Here we give a handful of specific examples of how very sensitive detectors developed for neutrino physics have been applied elsewhere, and vice versa. 

\begin{itemize}
\item Over more than a decade, a new germanium diode detector geometry, the P-type Point Contact (PPC) detector, has been developed to enable high energy resolution and high efficiency combined with large detector mass. This is achieved by careful design of internal electric fields and gradient impurities within the Ge. Detection of low energy recoils from the CEvNS process were a strong motivation for this development effort ({\it e.g.} ~\cite{CoGeNT:2010ols}). This technology forms the basis for 'Broad Energy Germaninum' (BEGe) detectors that are now a standard commercially available product~\cite{BEGe}.  The very good energy resolution over a wide energy range and high efficiency of the BEGe geometry is advantageous for many nuclear counting and spectroscopy applications. 

\item The basis for the Project8 direct neutrino mass measurement experiment is the newly invented technique of Cyclotron Radiation Emission Spectroscopy (CRES)~\cite{Project8:2014ivu}. Here, the energy of a single beta-decay electron is determined with very high resolution from its cyclotron radio frequency emissions as it propagates in a strong magnetic field. This technique is already being adapted for high-precision beta-decay shape measurements for nuclear physics studies~\cite{he6} and as a high resolution x-ray spectroscopy technique with application to nuclear nonproliferation~\cite{Kazkaz:2019umq}.

\end{itemize}

By way of contrast, very sensitive detectors primarily developed for applications or other areas of science can enable new neutrino physics experiments, with a handful described here. 

\begin{itemize}

\item Superconducting tunnel junctions (STJs) are cryogenic radiation detectors with ultra-high energy resolution for measurements of low-energy radiation. Initially developed for X-ray astrophysics and high-resolution soft X-ray spectroscopy in biophysics and material science, STJs are increasingly also used in nuclear and particle physics. This includes the characterization of nuclear recoils and low-energy transitions, {\it e.g.} for the development of nuclear clocks~\cite{Ponce:2018mnc}. STJs also enable the search for physics beyond the standard model, {\it e.g.} the BeEST sterile neutrino search~\cite{BeEST:2021loi,Friedrich:2020nze}. Among cryogenic detector technologies, STJs are distinguished by their relatively high count rate capabilities of several 1000 counts/s per pixel.

\item Magnetic microcalorimeters (MMCs), also known as metallic magnetic calorimeters, are cryogenic detectors whose ultra-high energy resolution is based on measuring changes in magnetization due to the absorption of radiation. They have been used for high-resolution X-ray, gamma-ray and alpha spectroscopy. MMCs have also been adapted for decay energy spectroscopy by fully embedding isotopes of interest  into the detector, {\it e.g.} to measure the neutrino mass~\cite{Mantegazzini:2021yed} and to search for the neutrino-less double beta decay~\cite{Alenkov:2019jis}. Among cryogenic detector technologies, MMCs are known for relatively fast detector response with large detector materials~\cite{Kim:2017xrs}, which makes it suitable for low background dark matter and neutrino experiments via phonon pulse shape discrimination technique~\cite{Kim:2020gni, MMC-cevns:2021loi}.

\end{itemize}

\subsection{Direct Application of Neutrinos to Nuclear Security and Nuclear Energy}
\label{sec:neutrinoApplications}

%[\textbf{Pending update:} summarize monitoring concepts and potential neutrino-based capabilities following the treatment of Ref.~\cite{Bernstein:2019hix} ]

There is an active community primarily comprised of neutrino physicists and nuclear engineers that seeks to explore direct use of reactor neutrino detection in nuclear security and nuclear energy. The regular Applied Antineutrino Physics (AAP) Workshop series~\cite{aap,aap2019,Bowden:2016ntq,Bernstein:2019cwx,Bergevin:2019tcg}, held since 2003, has been a forum for discussion of capability demonstrations, technical developments, use case studies, and occasional interaction with experts from the nonproliferation community. 

A comprehensive review of use of neutrinos for nuclear security applications can be found in Ref.~\cite{Bernstein:2019hix}. Interest in such applications derives from the fact that electron antineutrinos are an inevitable by-product of nuclear fission via beta decay of fission daughter products or neutron activation of stable nuclei found in the structure of nuclear energy systems. Neutrinos are a potentially appealing fission signature since they will be emitted from a facility effectively unattenuated, which allows non-intrusive monitoring via a detection system placed some distance away with no direct connection. Conversely, this same property makes neutrino detection difficult, especially at larger standoff (baseline) distances which places significant practical constraints on monitoring capabilities.     

Following the treatment of Ref.~\cite{Bernstein:2019hix}, there are three primary areas where neutrino-based approaches might provide a technical capability: fissile material production monitoring, reactor discovery and exclusion, and monitoring of spent fuel and reprocessing waste. These are described briefly here in a technical context, while an end-user perspective on the utility of these approaches is discussed in Section~\ref{sec:nuclear-applications}.

\subsubsection{Fissile material production monitoring} 

Besides the expected proportional variation with power, the total flux and energy spectrum of neutrinos emitted by a reactor also varies with fuel composition (see Ref.~\cite{Bernstein:2019hix} and references therein). This is because the fission of different isotopes populates different daughter product distributions. 
Neutrino observables like rate, energy spectrum, and the time evolution of those quantities can therefore provide information on the consumption and production of fissile material in a reactor. It is important to note that neutrino measurements represent an average over the reactor core, whereas conventional nuclear safeguards procedures are largely performed on discrete items like fuel assemblies. There are many studies that asses the ability of neutrino measurements to access quantities relevant to monitoring applications, many of which are summarized in Ref.~\cite{Bernstein:2019hix}. These include examples of verifying expected operations, distinguishing different operating modes, and sensitivity to changes in reactor fuel composition. 

Neutrino physics and application-oriented experiments have demonstrated relevant detection capabilities, as also summarized in Ref.~\cite{Bernstein:2019hix}. These include ``near-field'' demonstrations using ton-scale detectors within 10s of meters of a reactor (\textit{e.g.}~\cite{Kuvshinnikov:1990ry,Declais:1994su,Bowden:2006hu,NUCIFER:2015hdd,NEOS:2016wee,PROSPECT:2018dtt,Haghighat:2018mve,DANSS:2018fnn}) and ``mid-field'' demonstrations using 10 ton scale detectors within 100-1000~m of reactor complexes (\textit{e.g.}~\cite{Boehm:2001ik,CHOOZ:2002qts,DoubleChooz:2011ymz,DayaBay:2012fng,RENO:2012mkc}). Recent demonstrations of near-field detection without overburden add a potentially important capability that allows more flexibility in system placement for monitoring applications~\cite{PROSPECT:2018dtt,Haghighat:2018mve}. 

\subsubsection{Reactor discovery and exclusion} 

The ability to travel long distances without interacting allows the possibility to detect neutrinos emitted from a reactor at large standoff, with KamLAND representing a demonstration of this capability~\cite{KamLAND:2013rgu} . It follows that a neutrino detector could provide the capability to discover the operation of an unknown reactor within a surrounding geographical area, or conversely exclude the presence of a such a facility. Verifying the absence of undeclared facilities is a component of international safeguards agreements. The primary challenges for discovery and exclusion concepts are the large required detector size, the background antineutrino flux from known operating reactors, and cosmogenic backgrounds that can only be reduced by underground deployment.

The ability to discover the presence of a 50~MW$_{th}$ reactor within a dwell time of 1~year has been used as a scenario for describing the sensitivity of ``far-field'' discovery and exclusion concepts. Estimates presented in Ref.~\cite{Bernstein:2019hix} indicate that under favorable  background conditions (no cosmogenic background and low reactor background typical of the Southern Hemisphere), the presence of a 50~MW$_{th}$ reactor could be established ($3\sigma$) within 1 year using a water-based detector with 1,000~kt fiducial mass at a range of 200~km. Under less favorable background conditions (no cosmogenics, but reactor backgrounds consistent with Central Europe), that same very large detector would have similar sensitivity out to 50~km standoff. 
Recent work studying Gd-doped water Cherenkov detectors incorporating cosmogenic backgrounds and instrumental effects indicates that the practical range limit for this approach varies between $20-35$~km depending on the reactor background conditions~\cite{Li:2022tbb}. This is achieved with a detector size of $\approx50$~kt ($25$~kt fiducial), and beyond that scale there is little gain in sensitivity, absent further detector R\&D. For context, the area excluded by a 200~km radius is comparable to that of the U.S. state of Pennsylvania, while that excluded by a 35~km radius is comparable to the area of Rhode Island. 

\subsubsection{Monitoring of spent fuel and reprocessing waste}

The potential use of neutrinos for monitoring and quantification of spent nuclear fuel (SNF) and reprocessing waste (RW) is described in detail in Refs~\cite{Brdar:2016swo} and ~\cite{Bernstein:2019hix}. Both waste forms are highly radioactive and since the bulk of that activity occurs via beta decay they are neutrino sources. However, there are only a limited number of fission daughter products that generate neutrinos above the inverse beta decay threshold with an associated half-life relevant to long term SNF and RW monitoring, the most useful of which is the $^{90}$Sr/Y decay chain. Detecting neutrinos from SNF in dry cask storage for use in inventory verification appears possible, but very technically challenging due to low signal rates and the need for very good background rejection. Similar considerations apply to the potential use of neutrinos to quantify the amount or locations of RW. Large underground detectors at the 100-1000 ton scale could potentially be used to verify inventory of long-term geological repositories for SNF.

% This is an example of a figure, Fig.~\ref{fig:disappearance}.

% \begin{figure}[h!]
%     \centering
%     \includegraphics[height=3.5in,width=0.43\textwidth]{graphics/disappearance}
%     \includegraphics[height=3.7in,width=0.47\textwidth]{graphics/ComboLimit99_prelim.pdf}
%     \caption{Left panel: Comparison of present exclusion limits from various experiments obtained through searches for disappearance of muon neutrinos into sterile species assuming a 3+1 model.  The Gariazzo et al. region represents a global fit to neutrino oscillation data~\cite{Gariazzo_2015}. 
%     Right panel: The combined results of the disappearance measurements from \minosplus, \dayabay, and \bugey, compared to the appearance measurements from \lsnd and \miniboone.}
%     \label{fig:disappearance}
% \end{figure}

%%%%%%%%%%%%%%%%%%%%%%%%%%%%%%%%%%%%%%%%%%%%%%%%%%%%%%%%%%%
%\cleardoublepage

\FloatBarrier
\section{Nuclear Security and Nuclear Energy Applications -- An  End-User Perspective}
\label{sec:nuclear-applications}

\subsection{The Nu Tools Study}

The Nu Tools study~\cite{nuToolsSite} was commissioned by the DOE National Nuclear Security Administration (NNSA) Office of Defence Nuclear Nonproliferation Research and Development (DNN R\&D) to provide a user-derived perspective on the potential utility of neutrino detection for nuclear security and nuclear energy applications. Since most earlier studies have focused on characterizing neutrino signals and  detection technologies, one important goal of this utility-focused study was adding practical context that could help to guide current and future technology development. The major product of the study is the Nu Tools Final Report~\cite{Akindele:2021sbh}, which has been submitted as a contribution to Snowmass and is summarized here. 

The Nu Tools Study was conducted by a 14 person Executive group whose membership comprised a mixture of nuclear engineers and physicists drawn from academia and U.S. government laboratories. Input to the study was obtained by broad engagement with relevant user communities. This was accomplished through interviews with 41 experts with expertise in Nuclear Security and Safeguards, Reactor Design and Engineering, and Neutrino Physics and Technology, as well as a dedicated mini-workshop. Consistent with the focus on utility and end-user input, interviewees were selected with an emphasis on experts outside the physics research community, including international and domestic safeguards practitioners, nuclear reactor vendors and operators, and nuclear policy experts with experience in government agencies and non-governmental organizations.

Through synthesis of the engagement outcomes, the Nu Tools study developed a number of \textit{Cross Cutting Findings} that apply across all potential capabilities and use cases. These findings reveal features of end user thought and practice that will be important to incorporate into technology development efforts if they are be responsive to user need and move towards real world use. Brief summaries of the Cross Cutting Findings, adapted from the full text of Ref.~\cite{Akindele:2021sbh}, are given here. More detail explaining each finding can be found in that document.

\begin{itemize}
    \item \textbf{End User Engagement.} \textit{The neutrino technology R\&D community is only beginning to engage attentively with end-users, and further coordinated exchange is necessary to explore and develop potential use cases.} 
    
    Some end users see potential in neutrino technologies, but there is not a compelling use case that justifies their adoption at this time. 
To date, potential use cases have mostly been developed in the technical community with a focus on signals and detection, an approach that has not established enduring end-user connection or credibility. 
Systematic and sustained two-way exchange will be  necessary to develop mutual understanding between the technical and end user communities and identify use cases responsive to user needs.

    \item \textbf{Technical Readiness.} \textit{The incorporation of new technologies into the nuclear energy or security toolbox is a methodical process, requiring a novel system such as a neutrino detector to demonstrate sufficient technical readiness.} 
    
    The Technical Readiness Level (TRL) required by specific use cases will vary, but will generally be higher than currently demonstrated neutrino monitoring systems. 
    Safeguards end-users will generally expect demonstration at TRL 7-8, i.e. full-scale demonstration in operationally relevant environment. Demonstrations at this fidelity are important to demonstrate to users that a technology is sufficiently mature to provide the basis for safeguards determinations and provide user familiarity.
    As development of a neutrino technology proceeds through successively higher TRLs, end-user input should be an integral part of the process to ensure that it has features and performance required to meet user needs and implementation constraints.

    \item \textbf{Neutrino System Siting.} \textit{Siting of a neutrino-based system requires a balance between intrusiveness concerns and technical considerations, where the latter favor a siting as close as possible.}     
    
    Non-intrusiveness is viewed as a key advantage of neutrino-based monitoring approaches.
    Neutrino-based monitoring can assuage intrusiveness concerns, since no connection to facility process components is required to access a neutrino signal.
    Intrusiveness considerations could lead a cooperatively monitored party to prefer sites far from the reactor of interest. 
    However, a very strong impetus to negotiate the closest possible deployment site derives from implementation constraints related to neutrino detector size, cost, and construction timeline.

\end{itemize}

To evaluate the potential of use cases, the Nu Tools Utility Framework was developed. During the expert engagement process of the study, common themes emerged that were synthesized into the following four criteria:

\begin{enumerate}

\item \textbf{Capability need} for specific detection and/or monitoring capabilities\textit{is expressed by the end user community}. Such capabilities might not be presently available or they may be less effective than needed. Since cooperative nonproliferation activities are often conducted under tight resource constraints the 
expressed user need for a capability often considers the cost and effort associated with it. Potentially desirable capabilities may no longer meet user need beyond a certain cost, while on the other hand some capabilities may be perceived to have little  associated value even at modest cost. 

\item \textbf{Existence of a neutrino signal} \textit{is evaluated by the neutrino technology community} using well known physics and details of a specific use case.  

\item \textbf{Availability of a detector technology} \textit{is determined  by the neutrino technology community} considering whether a neutrino signal can be detected and backgrounds sufficiently controlled.  

\item \textbf{Implementation constraints} \textit{are expressed by the user community} and include factors like cost, workforce requirements (e.g. number of personnel and training), measurement timeliness, lead time to deployment, intrusiveness, and general logistical constraints. 

\end{enumerate}

Notably,  expertise on capability need and implementation constraints primarily resides in the nuclear security and nuclear engineering communities. Therefore engagement with these communities formed the basis for use case evaluation in the Nu Tools study. As described in the Cross-Cutting Findings, continued engagement of this type will continue to be important for advancing neutrino technologies from the laboratory to real world use. 

\subsection{Use Cases with Potential User Interest}

Applying the Utility Framework, the following use cases were found to have potential. All require further engagement, study, and definition, but can already provide important guidance to neutrino technology efforts. Brief summaries adapted from the full text of Ref.~\cite{Akindele:2021sbh} are given here. More detail explaining each finding can be found in that document.

\subsubsection{Advanced Reactor Safeguards}

\textit{Advanced reactors present novel safeguards challenges which represent possible use cases for neutrino monitoring.}

In contrast to safeguards for the current reactor fleet (described below), the Nu Tools engagement phase found a need being expressed for new safeguards methods being needed for some advanced reactor systems. In particular, advanced reactor concepts where the concept of item accountancy no longer applies were identified as an area where neutrino detection could provide a useful capability. There are many advanced reactor types under development, and the potential for a neutrino detector system to be developed largely independently since no physical connection to a facility is required may also be an attractive feature. Further detailed study is needed to understand parameters of interest and how well  neutrino detection can address these. Similar to other findings, end user implementation constraints like system cost remain a concern. 

\subsubsection{Future Nuclear Deals}

\textit{There is interest in the policy community in neutrino detection as a possible element of future nuclear deals.}

During the Nu Tools engagement phase, some experts expressed the view that neutrino detection warrants further consideration in the context of new treaties and agreements, with a greater likelihood of relevance for those involving a small number of countries. Reasons for this potential interest are that verification of new agreements may require novel capabilities and that a new treaty involving relatively few parties may have fewer constraints on verification approaches and be more open to negotiation compared to large established treaties. Also, agreements with relatively limited scope and a foundation in mutual confidence building are more amenable to the introduction of novel verification technologies. The possibility of combining verification functions with scientific cooperation between participating nations was highlighted as an attractive possibility by some experts.

Implementation constraints related to cost and technical readiness are also a concern in the context of future nuclear deals. As discussed above, considerations including cost favor siting as close as reasonably achievable to an individual reactor facility of interest, consistent with any  intrusiveness concerns being addressed.

\subsubsection{Spent Nuclear Fuel}

\textit{Non-destructive assay of dry casks is a capability need which could potentially be met by neutrino technology, whereas long-term geological repositories are unlikely to present a use case.}

The growing inventory of spent nuclear fuel in dry cask storage is primarily maintained under safeguards through a continuity of knowledge approach based on seals and video surveillance.  
The Nu Tools engagement phase found that non-destructive assay of dry cask contents is a presently unmet capability need, and neutrino signatures potentially contain sufficient information for this task. However, the weak neutrino emissions of spent fuel (relative to operating reactors) present a very significant detection challenge. 

The more permanent approach of spent nuclear fuel storage in geological repositories is still under development, as are approaches for safeguarding such facilities. In principle, neutrino detectors could supplement planned Containment and Surveillance methods. However, implementation challenges deriving from the low neutrino emissions and consequent requirement for large detectors likely outweigh the possible monitoring value.

\subsubsection{Post-Accident Response}

\textit{Determining the status of core assemblies and spent fuel is a capability need for post-accident response, but the applicability of neutrino detectors to these applications requires further study.}

The response  to the nuclear power accidents at Three Mile Island, Chernobyl, and Fukushima Daiichi was hampered by a lack of knowledge regarding the configuration and criticality status of the reactor fuel. The Nu Tools engagement process found that a capability need exists for instruments able to provide information of this type. In principle, neutrino detection could provide information on the presence of ongoing fission reactions in an accident scenario. However, little is presently known about fission power sensitivity requirements and any detector would have to operate in a very challenging post-accident environment. Further engagement and study will be required to better understand whether neutrino detection can provide a useful capability.  

\subsection{Use Cases with Limited User Interest}

Applying the Utility Framework, the following use cases, while technically possible in principle, were found to have limited end user interest. Brief summaries adapted from the full text of Ref.~\cite{Akindele:2021sbh} are given here. More detail explaining each finding can be found in that document.

\subsubsection{Current International Atomic Energy Agency (IAEA) Safeguards}

\textit{For the vast majority of reactors under current IAEA safeguards, the safeguards community is satisfied with the existing toolset and does not see a specific role for neutrinos.}

The current reactor safeguards approach largely relies on containment, surveillance, and item accountancy, and the Nu Tools engagement phase identified no significant capability gaps. These approaches suffice because the fuel in these reactors comes in the form of discrete, countable units. Interest from safeguards practitioners in new technologies is guided by operational ease and time savings without a significant increase in cost, which are criteria that current neutrino detection technologies do not meet.

Neutrino detectors can provide power and/or fuel burn-up measurements, but these quantities are rarely measured directly under current safeguards practice. Where measurements are made it is unclear that neutrino detectors would provide any benefits over existing tools, particularly given the current cost comparison. That is, neutrino detectors offer capabilities which are either a more expensive duplicate of existing capabilities, such as power measurements at research reactors, or which have no established role and are not seen as a need in the current IAEA practices, such as {\it in situ} burn-up measurements.

\subsubsection{Reactor Operations}

\textit{Utility of neutrino detectors as a component of instrumentation and control systems at existing reactors would be limited.}

Current reactor designs have accumulated many decades of operational experience and have a mature suite of instrumentation to guide operators, both in safety-related functions and overall performance monitoring.  Neutrino detection rates imply that the time required to determine a change in reactor state is long compared to what is needed for safety-critical instrumentation. Given these circumstances, the Nu Tools engagement phase identified no potential role for neutrinos as part of the operational infrastructure of nuclear reactors. 

%While neutrinos could provide cross calibration of instrumentation subject to harsh environments, e.g. a re-calibration reference for \BWR{}, no detailed studies of this potential capability have been performed to date.

\subsubsection{Non-Cooperative Reactor Monitoring or Discovery}

\textit{Implementation constraints related to required detector size, dwell time, distance, and backgrounds preclude consideration of neutrino detectors for non-cooperative monitoring.}

This finding applies across all detector size scales and operational ranges. Non-cooperative monitoring using smaller detectors at close-range is infeasible since  the required detector mass and dwell time are not compatible with a covert operation. 

Considering longer standoff distances, large neutrino detectors can in principle be used to discover an operating clandestine reactor within a radius of $\sim 100$~km, with the detector size at that range being similar to the largest under development by the neutrino physics community. 
An activity with the explicit goal of ``reactor discovery'' would be inherently non-cooperative, occurring from beyond the borders of a country of interest. This is because it is difficult to conceive of a host nation entering into a cooperative agreement that presumes the existence of undeclared facilities on its soil\footnote{Note that while the concept of ``reactor exclusion'' necessarily requires a reactor discovery capability, this is a goal distinct from discovery that is compatible with cooperative agreements.}. During the Nu Tools engagement phase, some experts expressed interest in such intelligence-gathering capabilities as a part of national technical means, e.g. to monitor states that do not grant access to international inspectors, while others considered this need fulfilled by existing capabilities like satellites. Similar considerations apply to the case of non-cooperative cross-border monitoring of a known facility.

However, general agreement was found that the practical challenges of long-range non-cooperative reactor monitoring outweigh the potential benefit it might provide. The major impediment is the required detector size to achieve sensitivity at long range, which  confronts multiple implementation constraints in the end-user community. These include construction and operation costs, construction timeline, and the potential political drawbacks resulting from a country conspicuously monitoring its neighbor. The combination of feasible range and the requirement for cross-border operation would also limit the locations such a capability could be applied to.

\section{Synergies with Particle Physics}
\label{sec:synergies}

Several Whitepapers~\cite{Akindele:2022dpr,Bernstein:2022aul,Alfonso:2022meh,Theia:2022uyh,Klein:2022tqr} and Letters of Interest~\cite{mutualbenefit:2021loi,RxPrediction:2021loi,Li6Organic:2021loi,workforce:2021loi,chandler:2021loi,prospect:2021loi,ROADSTR:2021loi,watchman:2021loi} submitted to the Snowmass process describe the mutual benefits that the neutrino physics and application communities could enjoy from strong and enduring engagement. In this section, the major topical areas in which synergies exist are discussed.

\subsection{Prediction of the Reactor Flux and Spectrum Source Term and Related Nuclear Data}
\label{subsec:sourceterm}

The topic which probably has the strongest scientific overlap between neutrino physics and direct neutrino applications is accurate prediction of reactor neutrino emissions. 
As discussed in detail in Ref.~\cite{Akindele:2022dpr} and the \textit{NF09 Neutrinos from Artificial Sources} Topical Group Report, there are persistent discrepancies between reactor flux and spectrum predictions and measurements. Resolution of these ``reactor anomalies'' is a very active multidisciplinary area of investigation. For example, a Letter of Interest on the topic~\cite{RxPrediction:2021loi} was endorsed by 17 collaborations and a number of individuals. 

The motivations for this activity are clear. Accurate reactor neutrino flux and spectrum predictions would have the following benefits:
\begin{itemize}
    \item Neutrino physics would benefit since the enormous reactor flux could continue to be a tool for discovery (Sections~\ref{sec:near-phys}, \ref{sec:cevns-phys}, \& \ref{sec:far-phys}).
    \item Neutrino applications would benefit since the ability to monitor and verify reactor operations and draw safeguards determinations will  depend upon having a reliable and trusted understanding of the flux and spectrum that should be emitted by a reactor under specified operating conditions.
    \item All communities would benefit from the improvements in underlying nuclear data and  modelling tools that will be be needed to produce accurate flux and spectrum predictions.
\end{itemize}

The steps required to improve our knowledge and ability to predict reactor flux and spectrum have been the subject of several dedicated meetings and workshops: 
\begin{itemize}
    \item The \textit{Technical Meeting on Nuclear Data for Anti-neutrino Spectra and Their Applications} was hosted by the IAEA Nuclear Data Section in 2019~\cite{bib:IAEA}.
    \item The \textit{Workshop on Nuclear Data for Reactor Antineutrino Measurements} (WoNDRAM) was held under the auspices of the Nuclear Data Inter-Agency Working Group (NDIAWG) in 2021~\cite{Romano:2022spd}.
\end{itemize}
The recommendations of these two activities have strong overlap with each other and Ref.~\cite{RxPrediction:2021loi}, suggesting a high degree of community consensus as to important steps to take. The highest priority recommendations related to reactor flux and spectrum of the most recent workshop, WoNDRAM~\cite{Romano:2022spd} are summarized here:
\begin{enumerate}
    \item Perform new electron spectra measurements following the neutron-induced fission of $^{235,238}$U and $^{239,241}$Pu.   
\item Perform correlated high-statistics antineutrino measurements at U.S.-based LEU and HEU reactors with a common high-precision detector; alternatively, perform an improved high-statistics $^{235}$U antineutrino flux and spectrum measurement at the accessible and high-power ORNL or NIST HEU research reactors.
 \item Perform an updated sensitivity study of the reactor source term to include fission yields as a function of reactor neutron spectra, fission product decay and capture cross sections, contributions from production and decay of minor actinides and capture on detector materials. 
\end{enumerate}

Recommendation (1) would provide modern data with which to perform  ``conversion'' predictions and provide an additional test of the hypothesis that the historical integral beta spectrum measurements of  $^{235}$U suffers from a normalization error~\cite{Akindele:2022dpr}. 
Recommendation (2) would provide benchmark neutrino spectra from the major fissile isotopes with strongly correlated systematic uncertainties for use as a direct reference in scientific and application studies and for validating other prediction approaches. 
Recommendation (3) would improve our understanding of the fidelity of reactor simulation inputs to neutrino flux and spectrum predictions (e.g. time evolution of fission daughter inventories) and provide guidance as to which underlying nuclear data could be improved.

Another notable recommendation that is consistent across  the IAEA and WoNDRAM meetings relates to \textit{	establishing standardized evaluations of reactor antineutrino datasets and predictions in a centralized, publicly accessible domain}~\cite{Romano:2022spd}. As is the case across all of particle physics, deliberate data standardization and stewardship is an essential best practice to enable long-term community access and to facilitate the rapid advancement of the field. Establishing data standards and an open software framework for flux and spectrum predictions by all commonly used methods would greatly facilitate progress by allowing meaningful inter-comparison between results, straightforward updates based on new data, and reduce barriers to entry for end-users from the particle physics, nuclear data, and neutrino applications communities. 

Several Snowmass contributions describe efforts that seek to measure reference neutrino spectra that would fully or partially address WoNDRAM recommendation (2):
\begin{itemize}
    \item JUNO-TAO will measure the time-evolving spectrum of a single commercial power reactor with unprecedented energy resolution~\cite{JunoTao:2021loi}.
    \item PROSPECT-II is a proposed upgrade to the PROSPECT experiment~\cite{Andriamirado:2022psq} based on a $^6$Li-doped liquid scintillator technology. Built as a relocatable detector, HEU (near pure $^{235}$U) and LEU spectra would be collected from the HFIR research reactor and a commercial power reactor respectively. Since this upgrade builds upon successful elements of PROSPECT it can proceed rapidly, with data collection commencing about 1 year after project start.  
    \item Two efforts are proposing readily mobile detector systems based on segmented plastic scintillator detector designs that incorporate $^6$Li (CHANDLER~\cite{chandler:2021loi} and ROADSTR~\cite{ROADSTR:2021loi}). A mobile capability of this type would clearly facilitate collection of systematics correlated reference spectra.
\end{itemize}

 Aside from the synergistic need for an accurate reactor flux and spectrum prediction capability, these measurement efforts  themselves represent a synergy between neutrino physics and neutrino applications,  as described in Snowmass LOI contributions from CHANDLER~\cite{chandler:2021loi}, PROSPECT~\cite{prospect:2021loi} and ROADSTR~\cite{ROADSTR:2021loi}. These measurements  occur in the ``near-field'' close to reactors and inherently demonstrate measurement capabilities relevant to applications. The recent demonstrations of above-ground reactor neutrino detection by PROSPECT~\cite{PROSPECT:2018dtt} and CHANDLER~\cite{Haghighat:2018mve} are especially notable in this regard, since they enable a much greater range of deployment locations and the concept of a readily mobile detection system that requires only utility support from a facility. The CHANDLER~\cite{chandler:2021loi} and ROADSTR~\cite{ROADSTR:2021loi} efforts have as an explicit goal development of application-appropriate readily mobile systems that can operate without overburden or other infrastructure support.

\subsection{Physics topics accessible using Near-field Inverse Beta Detectors}
\label{sec:near-phys}

Reactor neutrino detection in the near-field has seen overlap between physics experiments and reactor monitoring demonstrations for several decades. Current physics topics that motivate continued development of near-field technology include:

\begin{itemize}
     %\item \textit{NF03: Beyond Standard Model Physics:} 
     \item \textit{NF02: Understanding Experimental Neutrino Anomalies:} 
     Reactor short-baseline experiments are well suited to search for electron antineutrino disappearance consistent with the hints of a $\mathcal{O}(1\,{\rm eV})$ sterile neutrino~\cite{Abazajian:2012ys,Acero:2022wqg}.  This remains an important topic of research for ongoing detector projects such as PROSPECT~\cite{prospect:2021loi} and CHANDLER~\cite{chandler:2021loi}.  In particular, reactor sterile searches are enhanced by the use of segmented detectors, which is a specific near-field applications detector design choice that helps to suppress the surface-level background rates.  When paired with a compact research reactor, detector segmentation creates a range of baselines in a single, fixed detector module. 
    \item \textit{NF09 Neutrinos from Artificial Sources:} As discussed in Section~\ref{subsec:sourceterm}, knowledge of the reactor flux is critical to both applications and basic science.  Further, there are significant and persistent discrepancies between the measured reactor flux~\cite{Seo:2014xei} and theoretical models~\cite{Mueller:2011nm,Huber:2011wv}, in both the absolute rate and spectral shape.  Precision direct measurements of the reactor flux with a range of fuel compositions would inform the theoretical models and may highlight gaps in the nuclear data.  This work is best done with with high resolution detectors, operating in the near-field, where the effects of three neutrino mixing are negligible.   
\end{itemize}

The physics case for near-field detectors is discussed in detail in other reports~\cite{Akindele:2022dpr,Acero:2022wqg}. Current, planned, and proposed neutrino physics experiments at  relevant baselines with contributions to Snowmass include CHANDLER~\cite{chandler:2021loi}, PROSPECT~\cite{prospect:2021loi,Andriamirado:2022psq}, and LiquidO ~\cite{liquidO:2021loi}. These experimental efforts will continue to demonstrate important detection capabilities relevant to near-field applications across a range of target media and detector concepts.

\subsection{Physics topics accessible using CEvNS and Reactor Neutrinos}
\label{sec:cevns-phys}

While CEvNS detection, with its large cross section, holds  promise as a neutrino detection channel for future applications in reactor monitoring and nuclear security, in the near term, the R\&D needed to accomplish these far reaching goals has great overlap with the science goals of particle physics~\cite{Akindele:2022dpr,Abdullah:2022zue}.  These topics include:
\begin{itemize}
    \item \textit{NF02: Understanding Experimental Neutrino Anomalies:} CEvNS scattering could be used to perform a direct test of the sterile neutrino oscillation hypothesis through total neutral current disappearance.  
    \item \textit{NF03: Beyond Standard Model Physics:} The CEvNS scattering cross section and detected event rates are potentially sensitive to effects from non-standard interactions, neutrino magnetic moments and other electromagnetic properties.
    \item \textit{NF04 Neutrinos from Natural Sources:} Reactor CEvNS detectors would also be sensitive to neutrinos from other sources, including supernovae, geoneutrinos and solar neutrinos.
\end{itemize}

Current, planned, and proposed CEvNS experiments at reactors are also discussed in Refs.~\cite{Akindele:2022dpr,Abdullah:2022zue}. 

\subsection{Physics topics accessible using Mid \& Far-field Inverse Beta Decay Detectors}
\label{sec:far-phys}

The large underground neutrino detectors needed for Mid- and Far-field neutrino applications are also central to important physics topics of interest to the community~\cite{Bernstein:2022aul,Akindele:2022dpr}. 
Chooz\cite{CHOOZ:2002qts}, Palo Verde~\cite{Piepke:2002ju}, KamLAND~\cite{KamLAND:2013rgu}, Daya Bay~\cite{DayaBay:2019yxq}, Double Chooz~\cite{DoubleChooz:2019qbj}, and RENO~\cite{RENO:2020dxd} have measured reactor antineutrinos in the mid- and far-field thereby providing important capability demonstrations in addition to providing  fundamental knowledge of neutrino physics needed to consider neutrino applications. 

The  physics topics accessible with detectors at the size scale needed for mid- and far-field applications are discussed in more detail in the relevant Neutrino Frontier Topical Group Reports and several cases that use reactor neutrinos are summarized in Ref.~\cite{Akindele:2022dpr}. These topics include:
\begin{itemize}
    \item \textit{NF01 Neutrino Oscillations:} The physics of neutrino oscillations in the three-flavor paradigm, including mixing angles, the mass hierarchy, and CP-violation.
    \item \textit{NF04 Neutrinos from Natural Sources:} Neutrinos from solar, atmospheric, supernova, and geological sources being used to probe those objects and as a source for other physics goals.
    \item \textit{NF05 Neutrino Properties:} Searching for neutrinoless double beta decay.
\end{itemize}

Current, planned, and proposed neutrino physics experiments at the relevant size scale with contributions to Snowmass include Super-K~\cite{superk:2021loi} (Gd-doped water), SNO+~\cite{sno:2021loi} (water and liquid scintillator), Hyper-K~\cite{hyperk:2021loi} (water), JUNO~\cite{juno:2021loi} (liquid scintillator), \textsc{Theia}~\cite{Theia:2022uyh, theia:2021loi} (water-based liquid scintillator), and LiquidO ~\cite{liquidO:2021loi} (opaque liquid scintillator). In addition to the existing measurements listed above, these new experiments will continue to demonstrate important capabilities relevant to mid- and far-field applications across a range of target media, detector sizes, and standoff (baseline) distances. 

The potential DUNE ``Module of Opportunity''~\cite{MOO} represents a  connection between experiments of this type and  the U.S. long-baseline neutrino program. 
Large hydrogen-bearing liquid detectors would have distinct systematic uncertainties compared to Liquid Argon TPCs for the major long-baseline accelerator neutrino physics goals of DUNE. Additionally, they would enable complementary physics goals like some of those mentioned above. 

\subsection{Detector Technology Development Topics with Mutual Benefit to  Particle Physics and Nuclear Security}

Since neutrino detection lies at the heart of experimental neutrino physics and direct neutrino applications, it is natural that there are many synergies between these two endeavors. As already noted elsewhere, neutrino physics experiments have often developed neutrino detectors and performed neutrino measurements that serve as fundamental capability demonstrations for potential applications. As the field of neutrino applications has developed there are also opportunities for the converse to be true and for joint development activities.

Detailed description of the many ongoing detector technology development efforts is beyond the scope of this report, and can be found in that of the  \textit{NF10 Neutrino Detectors} Topical Group and associated white papers. Nonetheless, it is worthwhile to enumerate broad classes of detector development activities and the associated synergies between neutrino physics and applications.  

Historically, reactor antineutrino detection via Inverse Beta Decay has been based on scintillators and photo-detectors. Much work continues to expand and improve reactor antineutrino detection capabilities through development of new detection media, photon collection and sensing schemes, and detector geometries. Areas of active development are described in the \textit{NF10 Future Advances in Photon-Based Neutrino Detectors} white paper~\cite{Klein:2022tqr}, while some topics specific to reactor antineutrinos are summarized in Refs.~\cite{Akindele:2022dpr} and~\cite{Bernstein:2022aul}. Examples with contributions to Snowmass include:
\begin{itemize}
    \item Detection media: Water-based Liquid Scintillators~\cite{Theia:2022uyh,ait-wbls:2021loi,Klein:2022tqr}, scintillators with long-decay constants~\cite{Klein:2022tqr}, Gd-doping of water~\cite{ait-water:2021loi}, $^6$Li-doping of organic scintillators~\cite{Li6Organic:2021loi,ROADSTR:2021loi,Klein:2022tqr}, opaque scintillators~\cite{liquidO:2021loi,Klein:2022tqr}, and additively manufactured  scintillators~\cite{AMMS:2021loi, Klein:2022tqr}.
    
    \item Photon collection and sensing: Large Area Picosecond
Photodetectors (LAPPDs)~\cite{Klein:2022tqr}, Multi-wavelength-discriminating sensors such as dichroicons~\cite{Klein:2022tqr}, modular encapsulation of PMTs modules~\cite{pmtmodule:2021loi}, and wavelength shifting optical collectors~\cite{Bernstein:2022aul}. 
    
    \item Physically or virtually segmented geometries for improved event reconstruction, background rejection, and/or event directionality: CHANDLER~\cite{chandler:2021loi}, LiquidO~\cite{liquidO:2021loi,Klein:2022tqr}, NuLAT~\cite{nulat:2021loi}, PROSPECT~\cite{Andriamirado:2022psq,Klein:2022tqr}, and SANDD~\cite{SANDD:2021loi}.
\end{itemize}
These technologies can variously improve the detection performance, affordability, and deployability of photon-based reactor antineutrino detectors across all size scales. All of these parameters are of great mutual interest to efforts pursuing neutrino physics and neutrino application goals. As described in more detail in Sec.~\ref{sec:currentEfforts}, a suite of projects primarily supported by NNSA Defense Nuclear Nonproliferation R\&D  (the proposed Advanced Instrumentation Testbed (AIT)~\cite{watchman:2021loi,ait-water:2021loi,ait-wbls:2021loi,pmtmodule:2021loi}, \textsc{Eos}~\cite{Klein:2022tqr}, the BNL 30~ton Demonstrator~\cite{Klein:2022tqr}, and the Mobile Antineutrino Demonstrator), are noteworthy since they are developing technologies for neutrino applications that have strong synergies with the needs of neutrino physics experiments.

A newer area of detector development with potential relevance to neutrino applications is CE$\nu$NS detection. Motivated by  potential benefits of a higher interaction cross section and threshold-free interaction channel, CE$\nu$NS detection at reactors must overcome challenges related to the relatively small size of the single scattering energy deposition and potentially large environmental and instrumental backgrounds at low energies. Multiple approaches are being actively developed with the synergistic motivations of neutrino physics and applications. Detailed descriptions can be found in a dedicated  CE$\nu$NS whitepaper~\cite{cevns-snowmass} with summaries relevant to reactor antineutrino detection again given in Ref.~\cite{Akindele:2022dpr}. Examples with contributions to Snowmass include:
\begin{itemize}
\item Phonon-based detectors: Cryogenic bolometers (Mi$\nu$er)~\cite{MIvER:2021loi} and Magnetic Microcalorimeters~\cite{MMC-cevns:2021loi}
\item Noble liquid-based detectors: TPCs~\cite{nobleLiquid:2021loi} and bubble chambers~\cite{nobleBubble:2021loi}
\item Other approaches: Meta-stable water~\cite{metaWater:2021loi} and crystal damage measurements via color center production~\cite{Alfonso:2022meh}.
\end{itemize}

PALEOCCENE~\cite{Alfonso:2022meh}  is noteworthy for having introduced a passive detection concept that, if proven to be feasible, would be well suited to nuclear security inspection and verification activities. The ability to hand-carry and install a detection element with no power supply requirements would be highly valued by inspectors. That this detection concept also has potential neutrino and dark matter physics applications is yet another example of the strong synergies that exist between potential neutrino applications and these areas of physics that require very sensitive and selective detector technologies.

\subsection{Workforce Development for the Particle Physics and Nuclear Security Communities}

Training and mentoring a creative and diverse new generation of scientists and engineers is essential to the long-term success of the  High Energy Physics and nuclear security communities. Projects at the intersection of neutrino physics and neutrino applications have unique features that provide excellent training and education opportunities for early career community members while also adding a wider breadth of experience and opening new career pathways. 

For example, relatively small neutrino physics and applications oriented projects offer young scientists the  opportunity to be involved from the idea and design stage to data taking and analysis~\cite{workforce:2021loi,Akindele:2022dpr}. Experiments of order of 5 years duration offer invaluable training opportunities matched to the research timescale of postdocs (3 years) and graduate students (4-6 years). In addition, the relatively smaller size of collaborations of these experiments offer supportive and nurturing environments that are complementary to the opportunities found in large, international collaborations and reduce the threshold for early career scientists to get involved in multiple aspects of the projects.  

More broadly, neutrino applications offer a unique opportunity for collaborative engagement between applications-focused and neutrino science-focused community members~\cite{mutualbenefit:2021loi,Bernstein:2022aul,Akindele:2022dpr}.  
This will broaden the experience and expertise of early career community members, e.g. by providing an opportunity to learn about nuclear security policy and associated technical needs. 
This provides an additional non-traditional career pathway for neutrino physicists in US National Laboratories and industry. 
Conversely, projects that combine applications and science could be  be a way to introduce people from a broader set of backgrounds and interest to neutrino physics, helping to further diversify the field.
Investments from the National Nuclear Security Agency Office of Defence Nuclear Nonproliferation in Academic Consortia~\cite{NSSC,CNEC,MTV,ETI} are one example of how interaction between these communities can be facilitated.

\subsection{Facilities}

There are many facilities in the US that are being used for research that supports neutrino detection technology development and neutrino physics studies.  For CEvNS research the Spallation Neutron Source (SNS), at Oak Ridge National Laboratory~\cite{ORNL_nu:2021loi} (ORNL) has and will continue to be an important resource.  U.S. research reactors that are being used in detector development, flux measurements, and/or short baseline oscillation physics include the High Flux Isotope Reactor (HFIR) at ORNL~\cite{prospect:2021loi,ORNL_nu:2021loi}, and the Texas A\&M TRIGA reactor~\cite{MIvER:2021loi}. Research reactors at the National Institute of Standards and Technology (NIST) and the Idaho National Laboratory also have relevant characteristics and capabilities in this regard~\cite{PROSPECT:2015eri}. 
Many commercial reactors are also being used for these studies~\cite{chandler:2021loi,watchman:2021loi,Colaresi:2022obx}.
%\cleardoublepage

\section{Conclusions and Next Steps}
\label{sec:nextsteps}

From community discussion and submissions to Snowmass it is clear that technologies and workforce capabilities developed by the field of Neutrino Physics can have benefits in many other areas of science and society. This provides an additional motivation for societal investment in the field and pursue the scientific goals of the community. It is important that community members have such possibilities in mind and actively seek opportunities to highlight the broader impact of their research.

To enable continued progress on the development and adoption of neutrino applications will require coordinated support from government agencies, and engagement with end-user communities, as has been discussed in the Nu~Tools report~\cite{Akindele:2021sbh} and community submissions to Snowmass~\cite{mutualbenefit:2021loi, Akindele:2022dpr, Bernstein:2022aul}. 

\subsection{Inter-agency Coordination}
As was pointed out in Section~\ref{sec:synergies}, there is a great deal of overlap between the objectives of applications and security oriented agencies and and those of basic science.  So, it makes sense that these agencies should coordinate their funding to support technology development efforts that maximize the benefits to both science and applications.  Beyond that, DOE Office of Science and NSF funded researchers have valuable expertise on detector technologies and in calculating sensitivity to measurable quantities that may be relevant for applications.  Their guidance can be a real asset to program managers considering a role for neutrino detection in applications and security, even in the absence of an overlapping scientific motivation.  The Nu~Tools Study, which was commissioned by the National Nuclear Security Agency (NNSA) to explore practical roles for neutrinos in nuclear energy and security, is an excellent example of the neutrino science community providing guidance to a security agency. 

Going forward, we as the neutrino science community need to capitalize on the success of projects like the Nu~Tools study to strengthen the connections between all federal agencies interested in neutrino technologies and neutrino science, and  encourage greater coordination in the funding of projects that span the interface of science and applications.

%One recent example of Inter-agency coordination  is the Nuclear Data Inter-Agency Working Group (NDIAWG). Recognising the common needs and synergies between multiple programs, the NDIAWG brings together representatives from DOE Office of Science (Nuclear Physics and Isotope Production), DOE NNSA (Defence Nuclear Nonproliferation R\&D) and DOE Nuclear Energy, and research community representatives. The NDIAWG sponsors  the regular \textit{Workshop for Applied Nuclear Data Activities} (WANDA) series on nuclear data needs~\cite{wanda} and has previously released several Funding Opportunity Announcements.  

\subsection{End-user Engagement}
One of the cross-cutting finds of the Nu~Tools report identified coordinated and attentive end-user engagement as a critical step in the development of neutrino applications with real-world relevance.  The report goes on to note the dangers of poorly managed or haphazard engagement with a potential end-user community.  In particular, ``neutrino technology developers advocating for a specific approach without developing a deep understanding of the significant real-world goals and constraints," has lead to a wariness among some potential end-users and hindered collaboration with end-users on the exploration of some use cases.

Going forward, we as the neutrino science community would do well to conduct our end-user engagement with a coordinated approach, while recognizing that we do not yet possess a full understanding of the needs and constraints of potential end users. The Applied Antineutrino Physics (AAP) workshop series~\cite{aap} is one venue where this could occur, but opportunities to engage with end-users in forums more strongly associated with their own communities should also be pursued.

\subsection{Current Application-oriented Technology Development Efforts}
\label{sec:currentEfforts}

There are a number of ongoing efforts in application-oriented technology development with support from U.S. agencies outside the basic sciences. These are briefly described here to give the neutrino physics community an awareness of the breadth of such activities and their goals. In a handful of cases, these efforts also have support from basic science agencies thereby representing an ``existence proof'' of the possibility and value of such coordination.

The NNSA's Office for Defence Nuclear Nonproliferation R\&D (DNN R\&D) currently supports four neutrino detection technology research efforts. Three of these relate to technology development in support of large detectors for mid- and far-field concepts and are pursued with some degree of coordination with the Office of Science HEP Advanced Detector R\&D program:
\begin{itemize}
    \item The Advanced Instrumentation Testbed (AIT) is a proposed underground kiloton-scale detector testbed for reactor monitoring and
technology demonstrations~\cite{watchman:2021loi}. Possible target materials of interest for experiments at AIT are Gd-doped water~\cite{ait-water:2021loi} and water-based liquid scintillator~\cite{ait-wbls:2021loi}. 
%AIT is no longer planned for the Boubly site in the UK described in~\cite{watchman:2021loi} with a potential site at the Fairport mine in Ohio that previously hosted the IMB experiment being under consideration~\cite{Bernstein:2022aul}. 
    \item The \textsc{Eos} reconstruction and characterization demonstrator~\cite{Klein:2022tqr} is a 3-year program at LBNL that will test performance parameters of WbLS using a several ton-scale detector. The primary goal of \textsc{Eos} is to validate performance predictions for large-scale hybrid neutrino detectors by performing a data-driven demonstration of low-energy event reconstruction leveraging both Cherenkov and scintillation light simultaneously.
    \item The BNL 30~ton Deployment Demonstrator~\cite{Klein:2022tqr} is a 3 year program that will test large-scale deployment of different WbLS formulas without or with loading of metallic ions (i.e. Gd). The main goals are to retire risks in support of the potential deployment and operation of a kiloton-scale WbLS detectors by establishing the feasibility of and requirements for in-situ WbLS purification schemes, testing detector cleanliness and material compatibility, and confirming optical stability and light collection performance.
\end{itemize}

DNN R\&D has recently established the Mobile Antineutrino Demonstrator program. This 3 year effort was motivated in part  by the recent above-ground detection demonstrations by PROSPECT and CHANDLER, plus near-field use cases with potential end-user interest identified by the Nu~Tools Study. The goal of the program is to develop and construct a readily mobile ton-scale antineutrino detector system that requires no infrastructure support beyond electrical power and space for deployment. Requirements for the system will be defined with the \textit{End-User Engagement} and \textit{Technical Readiness} findings of the Nu~Tools Study in mind, so as to advance near-field systems of this type closer to real-world adoption. Detection technology options under consideration are the 3D segmented CHANDLER approach~\cite{chandler:2021loi} and 2D geometries using $^6$Li-doped PSD plastic scintillator~\cite{Klein:2022tqr} developed under the ROADSTR program~\cite{ROADSTR:2021loi}.

DNN R\&D also co-sponsors a number of nuclear data efforts seeking to improve reactor flux and spectrum predictions. These were selected and supported under the Nuclear Data Inter-Agency Working Group (NDIAWG) FOAs.

The Defense Advanced Research Projects Agency (DARPA) supports two efforts seeking to develop detectors capable of observing CEvNS at reactors. NCC-1701~\cite{Colaresi:2022obx,Akindele:2022dpr,Abdullah:2022zue} is based on mechanically cooled High Purity Ge diodes and has been deployed in a compact shielding package at a commercial reactor. The NCC-1701 effort also receives support from NSF~\cite{Colaresi:2022obx}.   NUXE~\cite{Akindele:2022dpr,Abdullah:2022zue} is based upon a low-threshold liquid xenon TPC. 

Upgrades to the MiniCHANDLER~\cite{chandler:2021loi,Haghighat:2018mve} prototype are being supported by the NSF Innovative Industrial Partnership program.
%\cleardoublepage

%\input{sections/conclusions}
%\cleardoublepage

\section{Acknowledgements}
\label{sec:acknowledgements}

The topical group would like to thank all the community members who contributed submissions to the Snowmass process and provided feedback on this report. 

This work was performed in part under the auspices of the U.S. Department of Energy by Lawrence Livermore National Laboratory under Contract DE-AC52-07NA27344. LLNL-TR-836071

%%%%%%%%%%%%%%%%%%%%%%%%%%%%%%%%%%%%%%%%%%%%%%%%%%%%%%%%%%%

%\clearpage

    % this is added just after end of document

% end stuff from init
%\cleardoublepage
%\renewcommand{\bibname}{References}
\renewcommand{\refname}{References}

%\bibliographystyle{apsrev4-1}
%\bibliographystyle{utphys}
% To understand the style chosen, see:
% https://arxiv.org/hypertex/bibstyles/ (very bottom -- additions) and
% https://www.sharelatex.com/learn/Bibtex_bibliography_styles
% July 2017, AH and BV (and AM)

%\begin{multicols}{2}[\printbibheading]
\bibliographystyle{JHEP}
\bibliography{common/main}
%\printbibliography
%\end{multicols}

\end{document}